\documentclass[conference]{IEEEtran}
\IEEEoverridecommandlockouts
\usepackage{cite}
\usepackage{amsmath,amssymb,amsfonts}
\usepackage{algorithmic}
\usepackage{graphicx}
\usepackage{textcomp}
\usepackage{xcolor}
\usepackage{subfig}

\def\BibTeX{{\rm B\kern-.05em{\sc i\kern-.025em b}\kern-.08em
    T\kern-.1667em\lower.7ex\hbox{E}\kern-.125emX}}
\begin{document}

\title{Deterministic Compressed Domain Analysis of Multi-channel ECG Measurements
}

\author{\IEEEauthorblockN{Dipayan Mitra}
\IEEEauthorblockA{\textit{Dept. of Systems and Computer Engineering} \\
\textit{Carleton University}\\
Ottawa, Canada \\
dipayan.mitra@carleton.ca}
\and
\IEEEauthorblockN{Sreeraman Rajan}
\IEEEauthorblockA{\textit{Dept. of Systems and Computer Engineering} \\
\textit{Carleton University}\\
Ottawa, Canada \\
sreeramanr@sce.carleton.ca}
\thanks{
  \textcopyright 2020 IEEE. Personal use of this material is permitted.  Permission from IEEE must be obtained for all other uses, in any current or future media, including reprinting/republishing this material for advertising or promotional purposes, creating new collective works, for resale or redistribution to servers or lists, or reuse of any copyrighted component of this work in other works.}}

\maketitle

\begin{abstract}
Continuous and long term acquisition of multi-channel ECG measurements are significant for diagnostic purposes. Compressive sensing has been proposed in the literature for obtaining continuous ECG measurements as it provides advantages including a reduced number of measurements, reduced power consumption and bandwidth for transmission. Reconstruction of the compressed ECG measurements is then done to analyze the measurements for diagnostic purposes. However, reconstruction of ECG measurements is computationally expensive. Therefore, in this paper, ECG analysis is carried out in the compressed domain without resorting to reconstruction. Multi-channel ECG measurements from MIT-BIH Arrhythmia database is used to validate the compressed domain ECG analysis.  ECG signals  are compressed at various compression ratios (CR) using morphology preserving deterministic sensing matrix. 
Structural similarity measures are used to quantitatively demonstrate the fidelity of the compressed measurements. R-peaks are detected in the compressed domain from compressed ECG measurements. Detection performance metrics such as sensitivity, positive predictivity and detection rate decrease as CR increases. 

\end{abstract}

\begin{IEEEkeywords}
ECG compression, wireless sensor network, compressive sensing, multi-channel ECG sensing, signal quality analysis.
\end{IEEEkeywords}

\section{Introduction}
Due to increase in illnesses like diabetes and hypertension, the risk of cardiovascular diseases (CVDs) has recently seen a sharp increase \cite{WHO}. As a consequence, the need for continuous acquisition and real time monitoring of vital signs, such as ECG, is needed for early detection of CVD so that early prevention modalities can be attempted. Internets of things (IoT) enabled Wearable sensor technology along with wireless body area networks (WBANs) offer continuous and real time ECG monitoring by continuously transmitting the acquired ECG measurements to a remote system for storage and further processing \cite{WBAN1, WBAN2, WBAN3}. A study by Kadrolkar \textit{et al.} showed that approximately $60\%$ of the small battery powered sensing devices' energy is consumed during data transmission for remote processing \cite{Kadrolkar}; hence data transmission may have to be done only on demand. As the wearable devices may not have adequate memory to store the acquired measurements, a reduction in the number of samples of the acquired measurements may also be in order. Compressive sensing offers a solution that reduces the amount of measurements and data transmission simultaneously. Thus compressive sensing has recently been proposed as signal acquisition technique and can ensure continuous sensing without the need for frequent battery replenishment. 

Compressive sensing (CS) acquires measurements far below the Nyquist rate and is recommended for compressible signals such EEG and ECG signals \cite{CS1},\cite{CS2},\cite{Mitra}. Although in literature some alternative compression algorithms have been proposed \cite{Ku}, CS based compression for ECG measurements have an edge over other techniques because of encoder design simplicity \cite{Yan, Polania}.  To effectively monitor cardiac activity, single channel ECG measurements often are segmented during acquisition through IoT enabled wearable sensors and recovery of compressed signal is outsourced. To recover the loss of quality due to segmentation during on-demand recovery, a Kronecker-based method was proposed in \cite{Hadi}, \cite{Mitra}. Also in \cite{Mitra_TIM}, as part of analysis of this improved recovery technique, deterministic and random sensing approaches were compared. Often ECG is  acquired on multiple channels when multiple leads are used (Holter monitoring). In such cases, although traditional one channel CS approaches may be acceptable, it would be inefficient. As the measurements acquired by multiple channels are generated by the electrical activity of heart, there exists a strong correlation between these measurements. Hence, these multi-channel and non-independent ECG measurements can be treated as jointly sparse. As a result, CS theory for multiple measurement vector (MMV), an extension of single measurement vector (SMV), may be an effective approach to acquire compressive ECG measurements continuously \cite{Ziniel, Mamaghanian}. 

Qiao \textit{et al.} used joint sparse model type 2 or JSM-2 to develop a two-step reconstruction scheme for jointly sparse measurements \cite{Qiao}. In \cite{Tigges}, compressive multiplexing approach was proposed for multi-channel ECG sensing and recovery. Sparse recovery algorithms exploiting joint sparsity of the ECG measurements, acquired by `resource-constrained' sensors, were shown in \cite{Mamaghanian_1, Sharma}. While the current state-of-the-art research  emphasizes the sparse recovery algorithm design, analysis on the jointly sparse multi-channel compressed ECG measurements (for the purpose of clinical evaluation) has not been widely reported.   

In this paper, we recommend and demonstrate compressed domain analysis of multi-channel ECG measurements without the need for recovery. A morphology preserving deterministic sensing model for acquiring multi-channel ECG measurements is presented in this paper. Deterministic sensing makes implementation in hardware easier, unlike random sensing matrices. As random sensing-based approach does not preserve morphology, it will require reconstruction of the compressively acquired signal for analysis. Unlike the existing methods in the literature the sensing technique presented in this paper does not need joint sparse recovery techniques for reconstruction.  In order to demonstrate the preservation of morphology, structural similarity is used in this work. The structural similarity of the measurements in the compressed domain is compared with that of the uncompressed measurements. A template-based correlation approach is used to quantify the structural similarity. The robustness of the sensing model is verified by detecting clinically significant feature, such as the R-peak from the QRS complex, in the compressed measurements. Performance evaluation metrics are used to provide a statistical analysis on pathologically significant ECG measurements chosen from MIT-BIH Arrhythmia database. 

The rest of the paper is organized as follows: in Section II background in CS has been discussed; CS based multi-channel ECG acquisition model has been presented in Section III; the quality evaluation metrics are introduced in Section IV; Section V contains the details of results and analysis; the paper ends in Section VI with a conclusion and scope of future work.

\textit{Notation:} In this work, boldfaced lower-case letters, e.g. \textbf{x}, denote vectors, whereas the boldfaced upper-case letters, e.g. \textbf{X}, denote matrices. Letter $n$ denotes index of the measurement and $[.]^T$ denotes the transpose operation. $cov(.)$ signifies covariance measure whereas $\sigma$ stands for standard deviation.   

\section{Background}
\subsection{CS for Single Channel system}
In classical CS theory, a $k$-sparse 1-D signal $\textbf{x}_N$ can be simultaneously sensed and compressed by a linear map onto $\textbf{y}_M$, also called measurement vector, for $M << N$, where,

\begin{equation}
    \textbf{y}_{M \times 1} = \mathbf{\Phi}_{M \times N} \textbf{x}_{N \times 1}
    \label{eq1}
\end{equation}

Here, $\mathbf{\Phi}$ is called the sensing matrix. Signal \textbf{x} is assumed to have a sparse representation in a basis, (sparsifying basis), $\mathbf{\Psi}$. Equation (\ref{eq1}) can be represented in the following form, 

\begin{equation}
    \textbf{y}_{M \times 1} = \mathbf{\Phi}_{M \times N} \mathbf{\Psi}_{N \times N} \textbf{s}_{N \times 1}
    \label{eq2}
\end{equation}
where \textbf{s} is the sparse vector.  Numerous algorithms have been developed to recover \textbf{x} from the compressed measurement vector \textbf{y} \cite{Recovery_1,Recovery_2,Recovery_4}. These are single measurement vector (SMV) recovery.

\subsection{CS for Multi-channel System}
Mathematically, a $t$-channel ECG measurement can be represented as,

\begin{equation}
    \textbf{X}_{N \times t} = [(\textbf{x}^1_N)^T, (\textbf{x}^2_N)^T, \dots,(\textbf{x}^t_N)^T]
    \label{eq3}
\end{equation}
where $\textbf{x}_N$ represents the ECG measurements acquired by individual channels and the superscripts (from $1 \dots t$) identify the respective channels. Similarly, $t$ measurement vectors can be obtained in the following way (forming MMV), 

\begin{equation}
    \textbf{Y}_{M \times t} = \mathbf{\Phi}_{M \times N} \mathbf{\Psi}_{N \times N} \textbf{S}_{N \times t}
    \label{eq4}
\end{equation}

Few recovery algorithms that exploit the temporal structure of the jointly sparse MMV are available \cite{Wipf, Hyder, Lee}.  Unfortunately, such algorithms are computationally expensive and cannot be carried out on a resource constrained wearable device.

 It was reported in \cite{Davenport_1, Davenport_2} that certain signal processing problems like, detection, classification, filtering can be performed on the compressed measurements itself, without the need to perform signal reconstruction.  However, the assumption was that compression was achieved using random $\mathbf{\Phi}_{M \times N}$.  In Section III we explore the idea of CS based morphology preserving jointly sparse ECG compression that may to lead to further signal processing in the compressed domain. 

\section{Deterministic Sensing for Multi-channel ECG Measurements}
Restricted isometry property (RIP) of $\mathbf{\Phi}$, introduced in \cite{Candes}, may be used for designing sensing matrices. 
Random matrices satisfy RIP condition and recovery can be guaranteed with overwhelming probability \cite{Baraniuk}. While random matrices offer a probabilistic notion in guaranteeing recovery of the compressed measurements, hardware realization for acquiring measurements is not straightforward. Hence, deterministic construction of the sensing matrices is recognized as a viable alternative \cite{Li}. In literature, several deterministic matrix construction techniques were proposed based on coding theory \cite{Deter_1, Deter_2, Deter_3}. In this work, we use linear filtering based deterministic binary block diagonal (DBBD) matrix as $\mathbf{\Phi}$, as it is easily implementable in a measurement system \cite{DBBD}. 

The DBBD sensing matrix can be viewed as a linear filter block followed by a decimation of $\frac{N}{M}$. %
Order of the filter and decimation is determined by CR. For example, for CR = $50\%$, sensing matrix would be a second order linear phase filter followed by decimation of factor $2$. Accordingly, appropriate representation can be found for higher CRs.

\section{Quality Evaluation Metrics}
To evaluate the quality of the compressed measurements, structural similarity and ability to detect fiduciary points are considered.  In this work, R peak is used as fiduciary point for evaluation purposes. 

\subsection{Structural Similarity}
In order to compare the structural similarity of the compressed measurements with that of the uncompressed ones, a block-based correlation approach was applied. Each compressed measurement was segmented into smaller blocks of length equal to individual beat length. Correlation between individual segments were obtained using Pearson's correlation coefficient (CC). Pearson's CC between two segments $A$ and $B$ is determined as follows,

\begin{equation}
    corr(A,B) = \frac{cov(A,B)}{\sigma_A \sigma_B}
\end{equation}
Note that $\sigma_A$ and $\sigma_B$ represented standard deviations of $A$ and $B$ respectively.

Pearson's CC of compressed and original measurements were compared to quantitatively infer about the structural similarity between both the measurements.

\subsection{Fiduciary Point Detection: R-peak Detection}
In order to demonstrate the ability to analyze ECG signal, R-peak detection is considered in this work.  Pan-Tompkins QRS detection algorithm is applied on the compressed ECG measurements \cite{Pan}. True R-peak is considered as detected, if the estimated location of the peak falls exactly on the peak of the QRS complex and is considered as true positive (TP).  If the estimated peak location is different from the actual peak location (obtained through visual inspection), then the peak is declared as false  and taken as false positive (FP). If no peak is detected when it should be, then it is declared as false negative (FN).  To evaluate the performance of QRS detection algorithm on the compressed measurements, the following metrics  in terms of number of TP, FP and FN were chosen for the evaluation purpose: sensitivity ($Se$), positive predictivity ($P+$), F measure ($F$) and detection error rate ($DER$). They are given below:
\begin{equation}
    Se (\%) = \frac{TP}{TP + FN} \times 100 \%
\end{equation}
\begin{equation}
    P+ (\%) = \frac{TP}{TP+FP} \times 100 \% 
\end{equation}
\begin{equation}
    F (\%) = \frac{2 \times TP}{2 \times TP + FN + FP} \times 100 \% 
\end{equation}
\begin{equation}
    DER (\%) = \frac{FP + FN}{TB} \times 100 \% 
\end{equation}
where TB stands for total number of beats.

\section{Result and Analysis}
To evaluate performance in compressed domain,  $18$ ECG measurements, representing `large variety of pathological cases', from MIT-Arrhythmia database were chosen \cite{MIT, WDD}. Two channel ECG measurements were digitized at a rate of 360 samples/second over 10 mV range and 11-bit resolution \cite{MIT_1}. In our analysis, $10240$ samples of measurements, for each channel, were compressed using DBBD matrix.   

\begin{figure}[htp]
\centering 
    \subfloat[]{\label{1}\includegraphics[width = 3.4in]{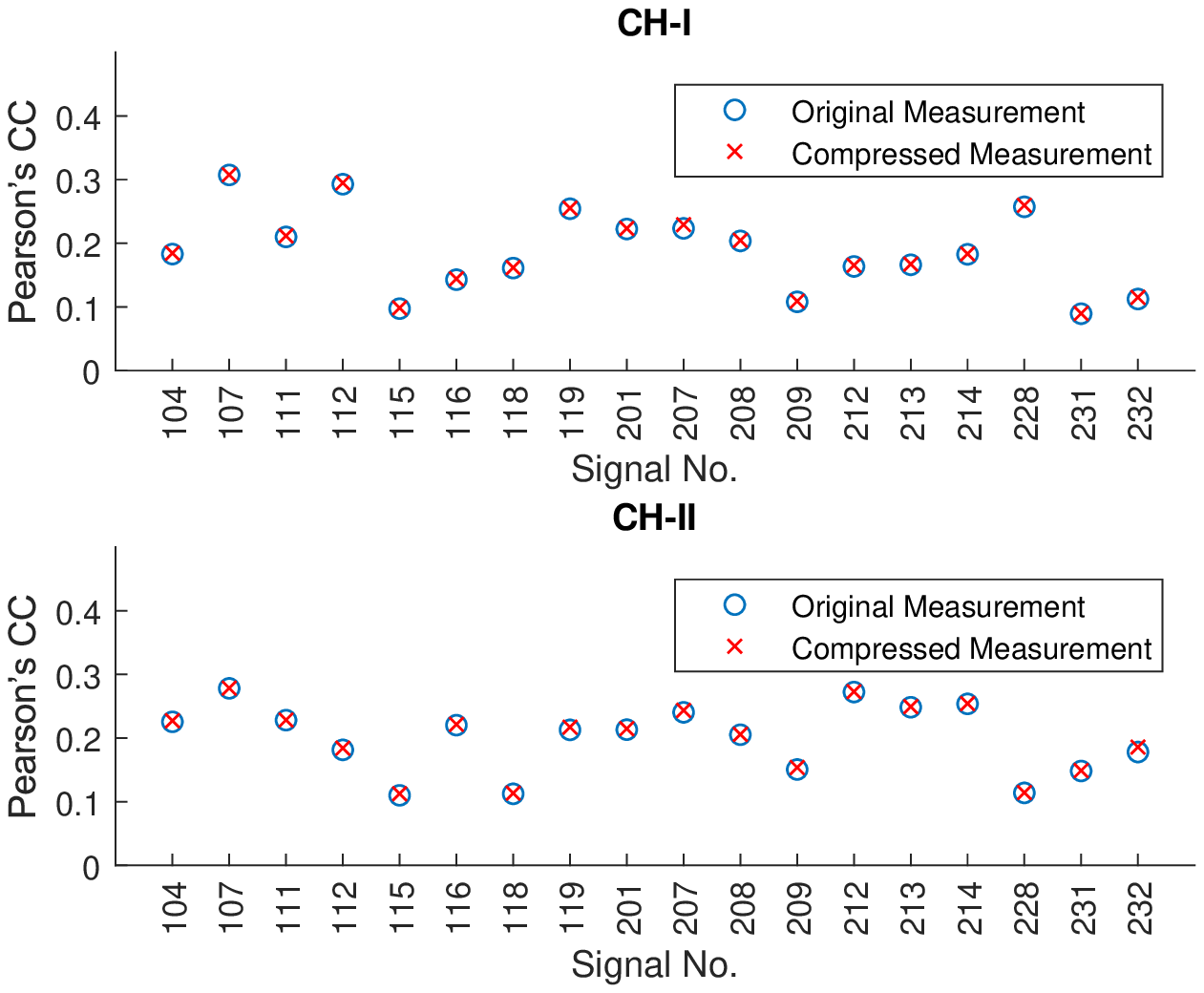}} \\
    \subfloat[]{\label{2}\includegraphics[width = 3.4in]{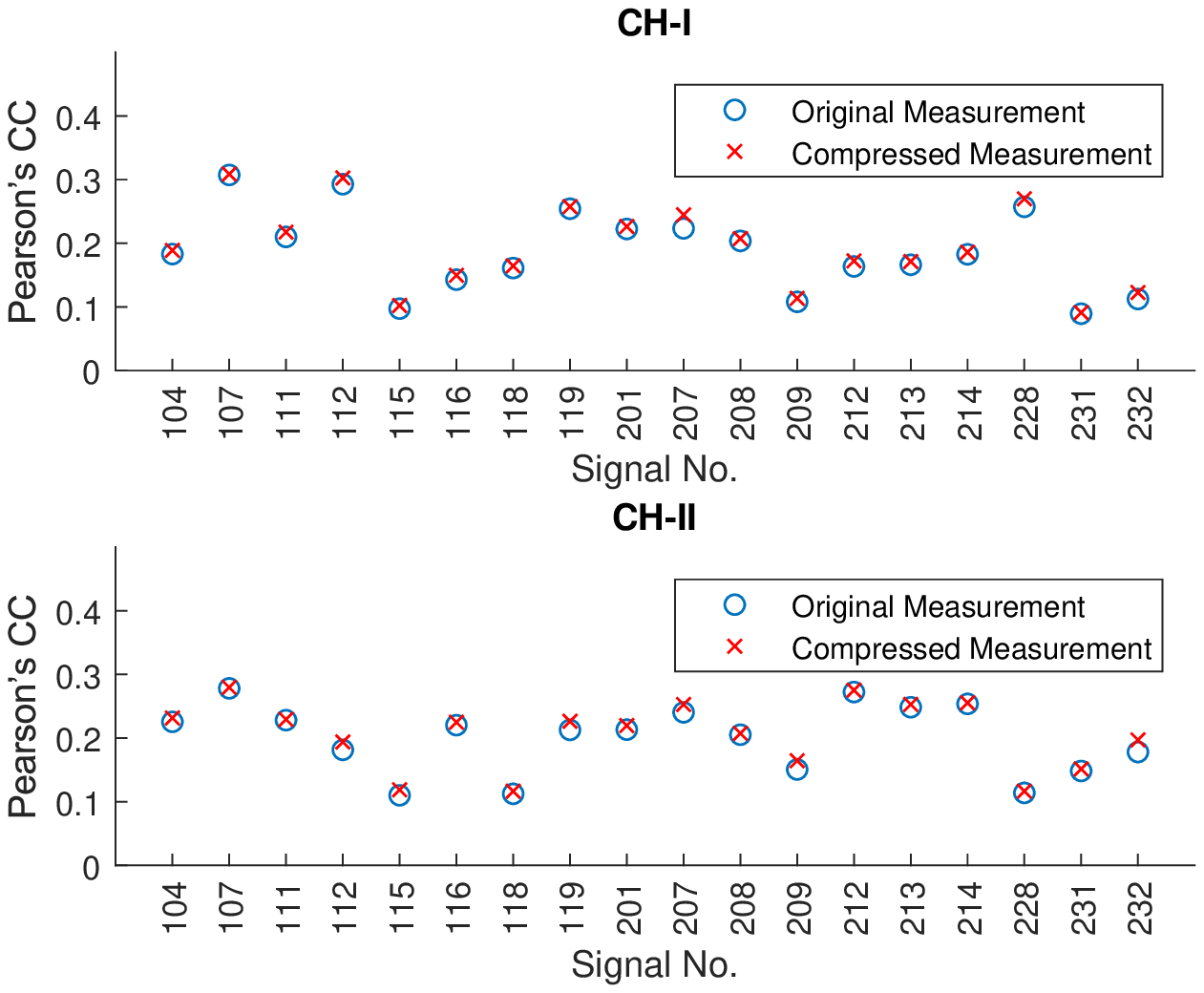}}\\
    \subfloat[]{\label{3}\includegraphics[width = 3.4in]{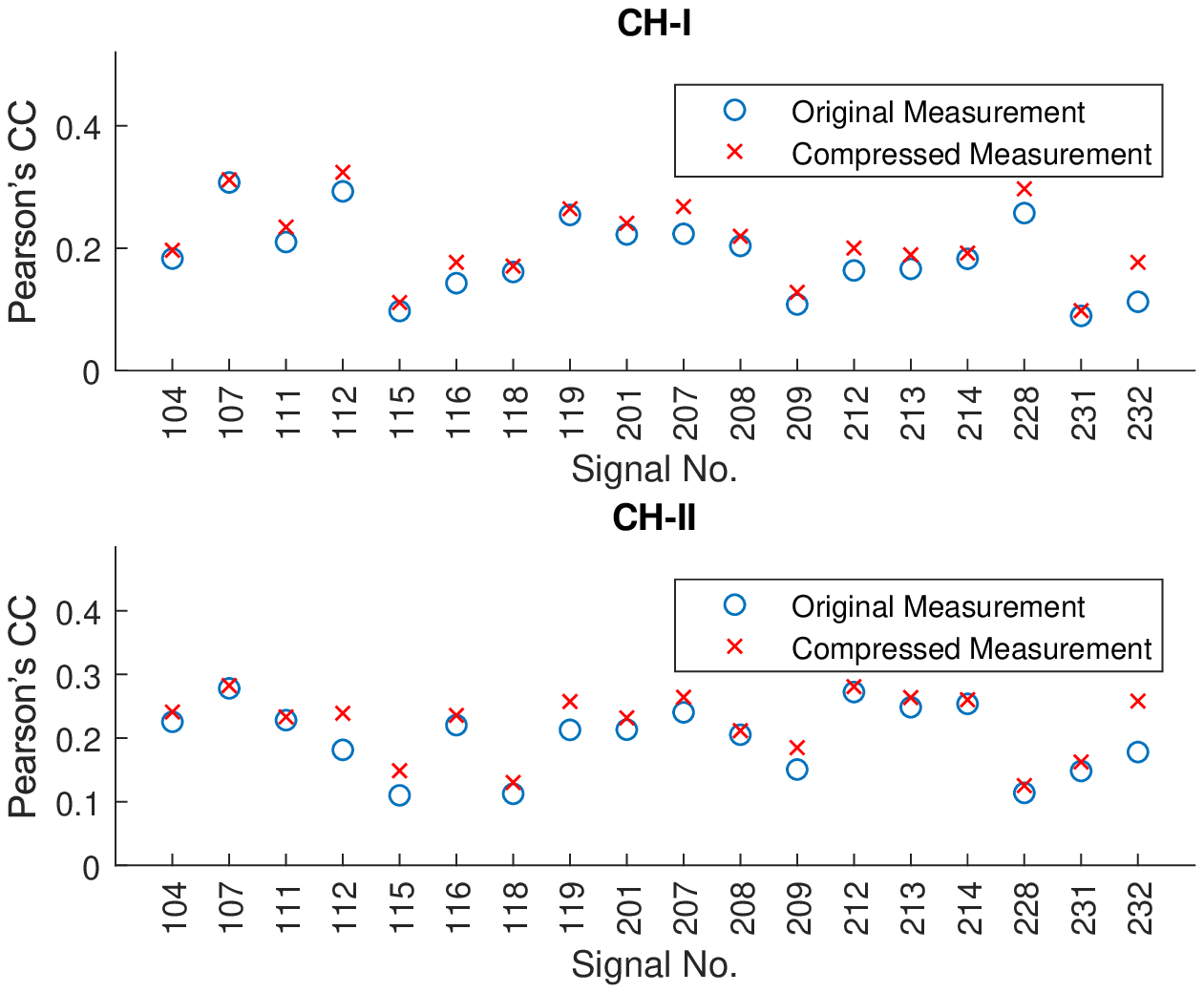}}
 \caption{Structural similarity analysis at (a) CR = $50\%$, (b) CR = $75\%$, (c) CR = $87.5\%$ (two channels are marked with \textbf{CH-I} and \textbf{CH-II}). }
    
    \label{fig3}
\end{figure}

To verify the structural similarity, while implementing the deterministic sensing model, we segmented the measurements (each of length $10240$) into $40$ smaller measurements (each of length $256$ for original measurement and of length $128$ or $64$ or $32$ compressed measurements with CRs = $50\%$ or $75\%$ or $87.5\%$ respectively). Segmentation was based on average beat length calculated by RR intervals (duration $\approx$ 0.2 sec) in the uncompressed domain. We refer these small segments of ECG measurements as \textit{templates}. Pearson's CC, described in Section-IV\textit{(A)}, was calculated for each of these \textit{templates} and an average estimate obtained for each measurement, which is used as a metric. Fig. \ref{fig3} shows the structural similarity between the original and compressed ECG measurements for different CRs (CR = $50\%$, $75\%$ and $87.5\%$) for both the channels. From analysis, it can be inferred that the deterministic sensing model preserves the morphology of the original measurements, which might be useful for application of signal processing algorithms on the compressed measurements itself (without further processing). However, for higher CRs, such as CR = $87.5\%$, slight degradation in similarity measure was observed. 

\begin{figure}[htp]
\centering 
    \subfloat[]{\label{11}\includegraphics[width =        3.2in, height = 1.6in]{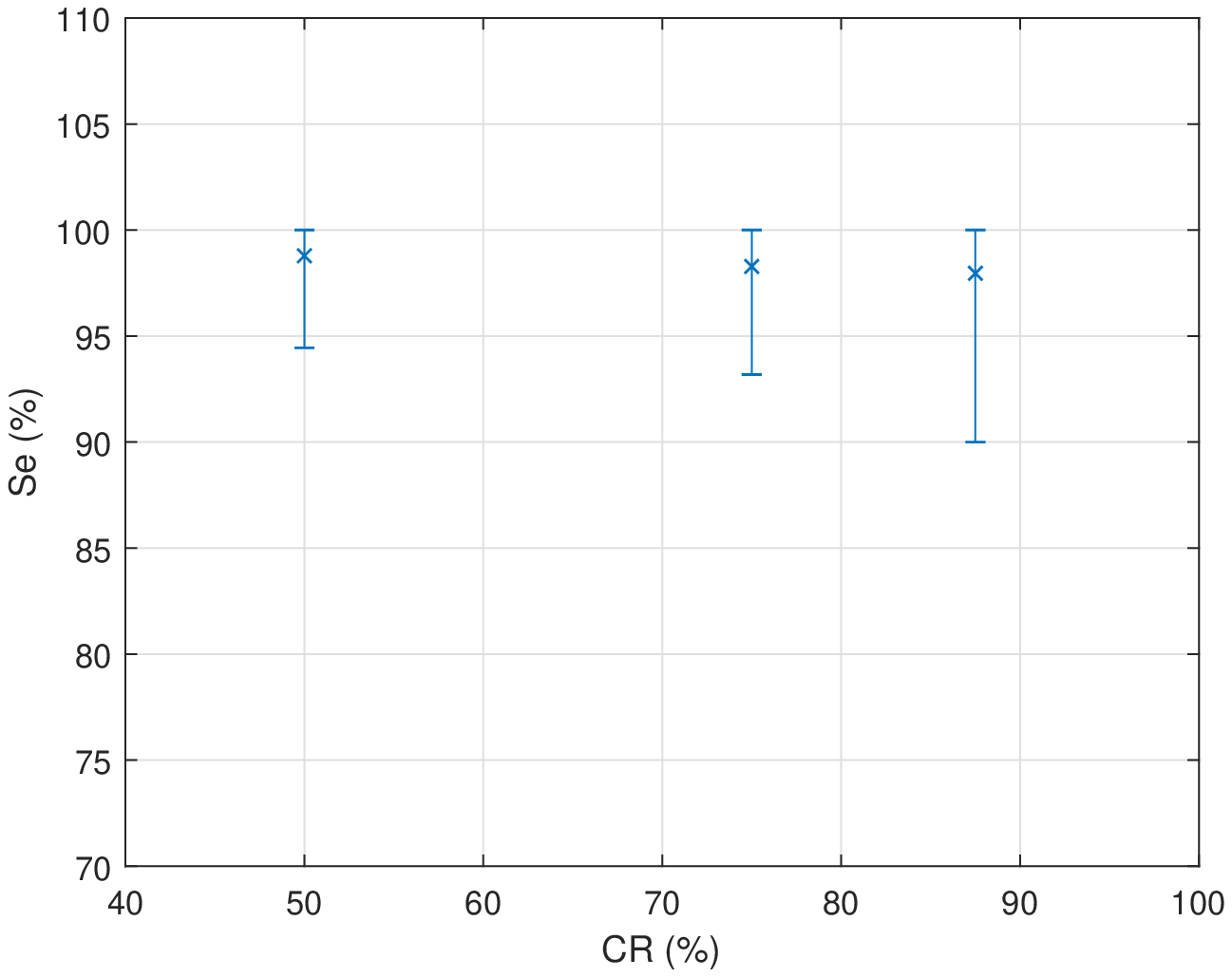}} \\
    \subfloat[]{\label{21}\includegraphics[width = 3.2in, height = 1.7in]{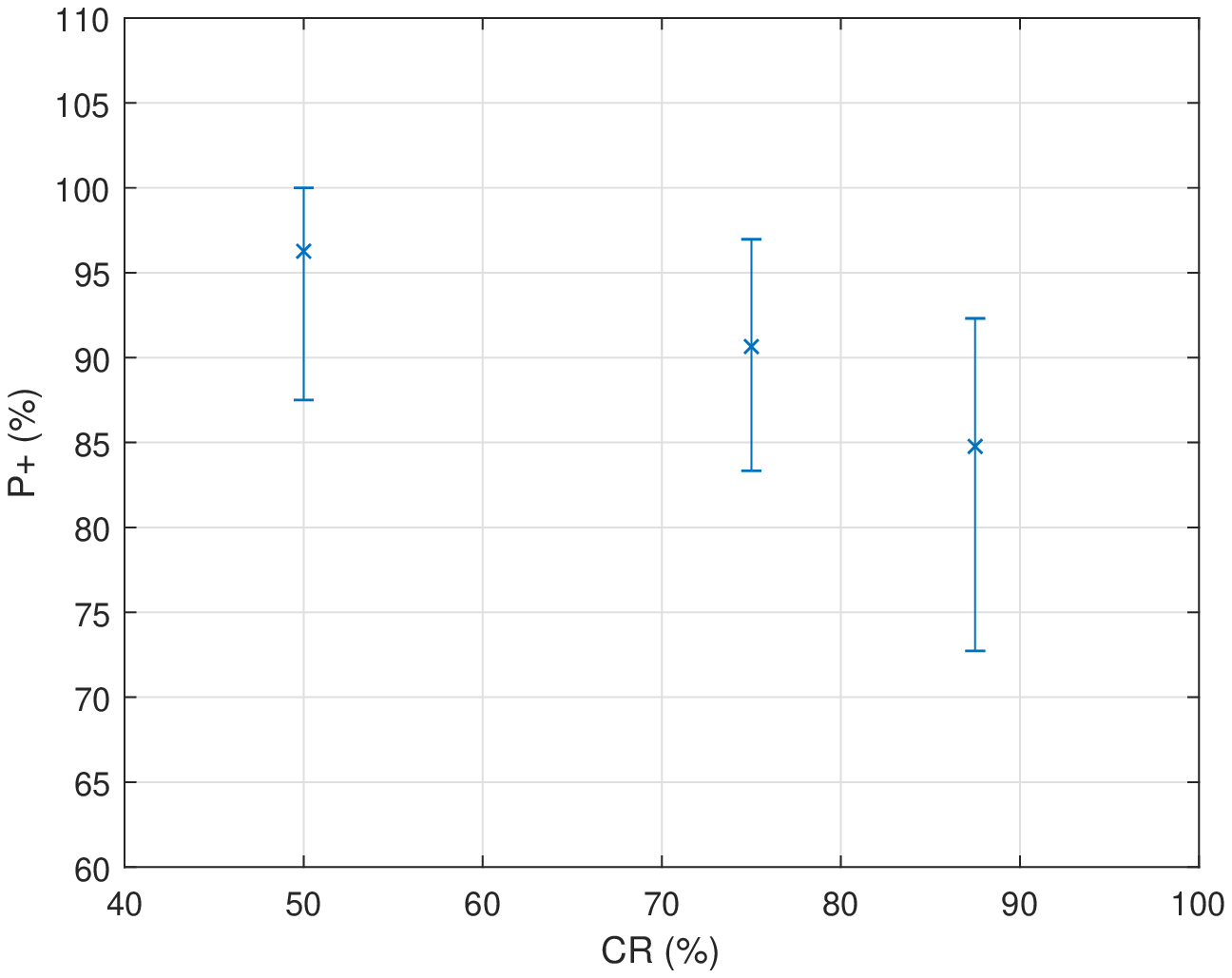}}\\
    \subfloat[]{\label{31}\includegraphics[width = 3.2in, height = 1.6in]{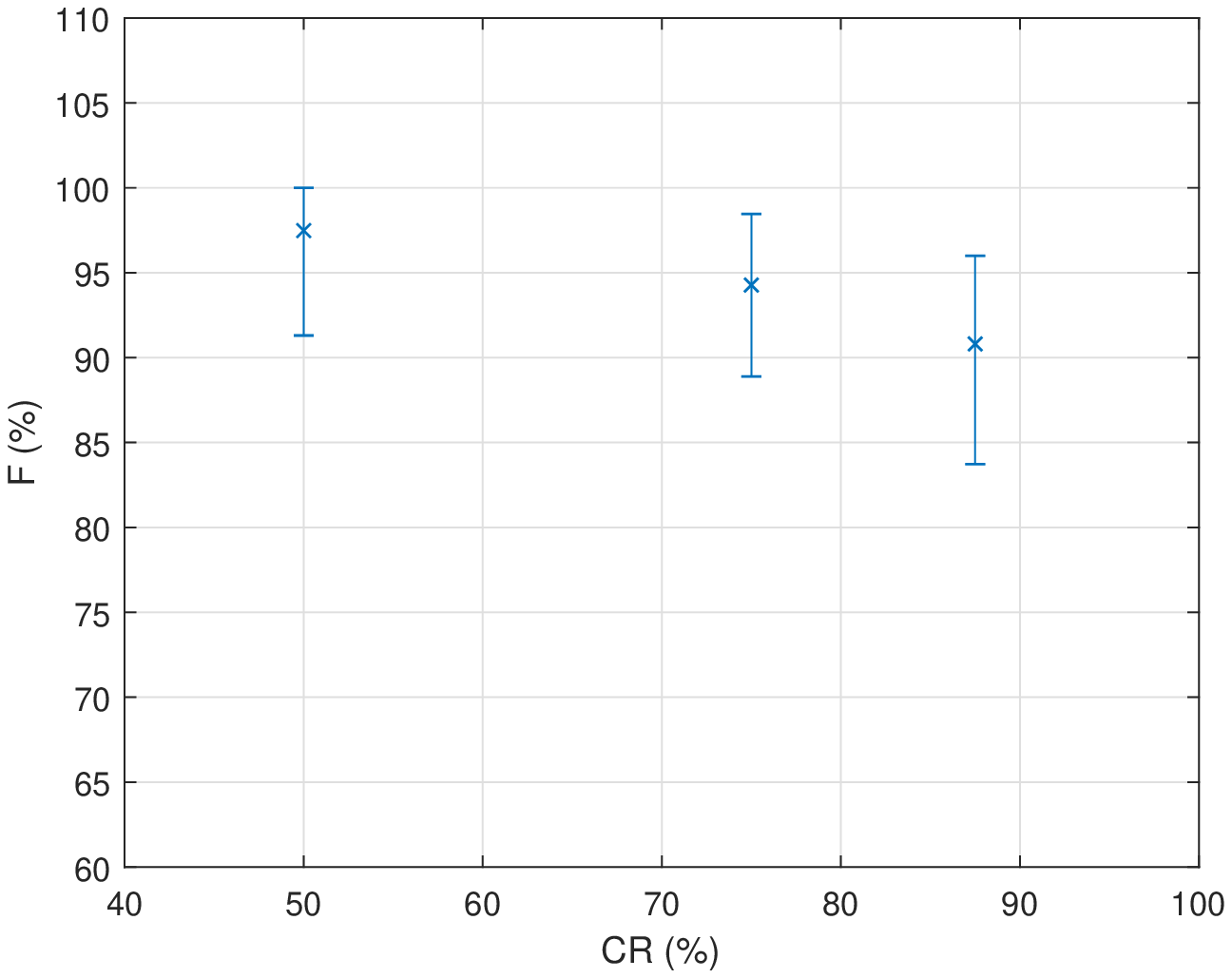}}\\
    \subfloat[]{\label{41}\includegraphics[width = 3.2in, height = 1.6in]{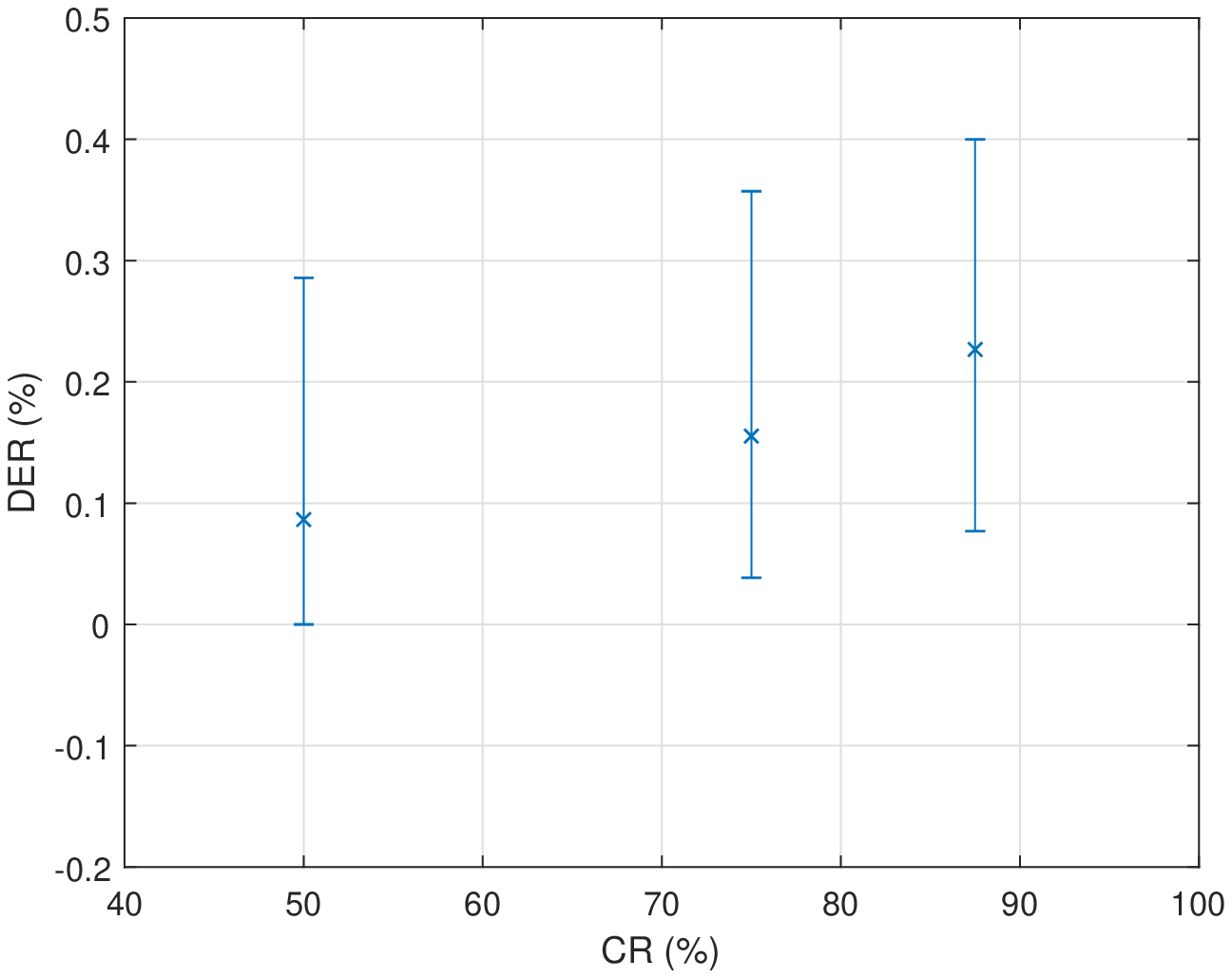}}\\
    \caption{Performance of R-peak detection on the compressed measurements (of CH-I) for varying CRs (CR = $50\%$, $75\%$ and $87.5\%$).  (a) Sensitivity analysis, (b) Positive predictivity analysis, (c) F-measure and (d) DER analysis.}
    \label{fig4}
\end{figure}

We carried out statistical analysis, mentioned in Section-IV\textit{(B)}, to verify the possibility of performing signal processing, like anomaly detection, on the compressed measurements. Fig. \ref{fig4} shows the analytical results, representing average statistical analysis on detection of R-peak from the compressed measurements (without designing any algorithm dedicated to compressed domain R-peak detection), for CRs = $50\%$, $75\%$ and $87.5\%$. Analysis was performed on individual compressed ECG measurements, for both channels. An average over all $18$ ECG measurements for channel-1 has been reported in this paper. Analysis for channel-2 has not been shown because of space constraint. Analytical results of the average values of sensitivity, positive predictivity and F-measure indicate that the clinically significant features of the ECG measurements remain unaffected using the deterministic sensing scheme. Low values of detection error rate signify the possibility of correct detection of R-peak, from the compressed measurements. From analysis, it is evident that with the increase of CR, detection performance decrease.   

\section{Conclusion}
In this paper, we presented a model for continuous multi-channel ECG sensing and analysis based on deterministic compressive sensing approach. Linear filtering based DBBD deterministic matrix, easily implementable in hardware, was used as sensing matrix. We formed \textit{templates} and used correlation between the compressed and the uncompressed ECG \textit{templates} to quantify the structural similarity between the two, for varying CRs. We also performed R-peak detection on the compressed ECG measurements and presented statistical analysis. The analysis was performed without designing or modifying any new algorithm for compressed domain processing. Moreover, the analysis was presented for varying CRs, by avoiding a computationally expensive joint sparse recovery for multi-channel ECG measurements. 

Although, the proposed sensing model preserves the morphology of the measurements in the compressed domain, for some applications measurements are required to be encrypted for privacy purposes. Future studies would include the application of anomaly detection using machine learning algorithms on the compressive measurements, sensed using random sensing, to preserve privacy. Effectiveness of the sensing model in presence of measurement noise would also be studied. Dimensionality reduction, offered by compressive sensing, would enable hardware based implementation of anomaly detection for continuous and long-term multi-channel ECG monitoring. 

\section{Acknowledgement}
The authors would like to acknowledge the financial support from NSERC.

\end{document}